\documentclass[12pt, english, twoside]{article} 

\usepackage[letterpaper]{geometry}
\usepackage[titletoc,title]{appendix}

\usepackage{fancyhdr} 
\fancypagestyle{plain} 
{
\fancyhead[LCR]{} 
\fancyhead[C]{}
     
}

\renewenvironment{abstract}
 {\small
  \begin{center}
\normalsize  \textnormal{ABSTRACT\\} \vspace{-0em}\vspace{0pt}
  \end{center}
  \list{}{%
    \setlength{\leftmargin}{0in}%
    \setlength{\rightmargin}{\leftmargin}%
  }%
  \item\relax}
 {\endlist}

\usepackage{sectsty}
\allsectionsfont{\large\raggedright\centering}

\usepackage[]{tocloft}
\addtocontents{toc}{\cftpagenumbersoff{section}} 
\addtocontents{toc}{\cftpagenumbersoff{subsection}} 

\makeatletter
\renewcommand{\@cftmaketoctitle}{}

\renewcommand{\@makefntext}[1]{%
  \setlength{\parindent}{0pt}%
  \begin{list}{}{\setlength{\labelwidth}{6mm}
    \setlength{\leftmargin}{\labelwidth}%
    \setlength{\labelsep}{5pt}
     \setlength{\itemsep}{0pt}%
      \setlength{\parsep}{0pt}%
      \setlength{\topsep}{-3pt}
    \footnotesize}%
  \item[\@textsuperscript{\@thefnmark}\hfil ]#1
  \end{list}%
}

\makeatother

\usepackage{authblk}

\usepackage{graphicx, listings, setspace, hyperref, verbatim}
\usepackage{amsmath, amssymb, mathtools, esint}
\usepackage{tikz, tabu, tabularx, makecell, gensymb, adjustbox, subcaption, float}
\usetikzlibrary{shapes, shadows, arrows}
\usepackage{textcomp}
\usepackage{mathptmx}

\usepackage[T1]{fontenc}
\usepackage[utf8]{inputenc}
\usepackage{mathptmx}

\newcommand\bs{\begin{singlespace}} 			
\newcommand\es{\end{singlespace}} 		
\newcommand\bq{\begin{quote}\begin{singlespace}\small}	
\newcommand\eq{\end{singlespace}\end{quote}}
\newcommand\be{\begin{equation}} 			
\newcommand\ee{\end{equation}}

\let\baraccent=\= 
\renewcommand{\=}[1]{\stackrel{#1}{=}} 

\hypersetup{
	pdftitle={De-Idealizing De-Idealization},
	pdfauthor={Yichen Luo and Eugene Y. S. Chua},
	pdfsubject={},
	pdfkeywords={},
	pdfcreator={XeLaTeX},
	pdfproducer={XeLaTeX},
	pdftoolbar=false,	
	pdfmenubar=true,	
	pdffitwindow=false,	
	pdfstartview={FitH},	
	unicode=true,		
	pdfnewwindow=true,	
	colorlinks=true,	
	linkcolor=black,	
	citecolor=black,	
	filecolor=black,	
	urlcolor=blue		
}

\usepackage{filecontents}

\usepackage[main=british]{babel}
\usepackage[babel=true]{csquotes} 

\usepackage[backend=biber, style=ext-authoryear-comp,  maxcitenames=2, dashed=false, url=false, doi=false, eprint=true, giveninits=true, innamebeforetitle=true, maxbibnames=6]{biblatex} 
\DeclareFieldFormat{labeldate}{\mkbibbrackets{#1}} 

\DeclareFieldFormat{postnote}{\mkpageprefix[pagination][\mkcomprange]{#1}} 

\AtEveryBibitem{%
 \clearfield{note}%
  \clearfield{series}%
 \clearfield{number}%
 \clearfield{pagetotal}%
 \clearfield{chapter}%
  }

\renewcommand{\intitlepunct}{\addspace\nopunct} 

\DeclareDelimFormat[bib,biblist]{nametitledelim}{\addcolon\space} 

\DeclareFieldFormat{biblabeldate}{\mkbibbrackets{#1}} 

\DeclareFieldFormat{editortype}{\mkbibparens{\textit{#1}}} 
\DeclareDelimFormat{editortypedelim}{\addspace}
\DeclareDelimFormat[bib,biblist]{innametitledelim}{\addcomma\space} 

\DeclareFieldFormat{eprint}{\printtext{available at}\addspace \textless#1\textgreater} 
  
\DeclareFieldFormat[article,periodical]{volume}{\textbf{#1}} 

\DeclareFieldFormat[inbook, incollection, book]{volume}{\bibcpstring{volume}~#1} 

\DeclareFieldFormat{pages}{\mkpageprefix[pagination][\mkcomprange]{#1}} 

\renewbibmacro{in:}{%
\ifboolexpr{%
test {\ifentrytype{article}}%
or
test {\ifentrytype{inproceedings}}%
}
{}
{\printtext{\bibstring{in}\intitlepunct}}
}

\usepackage{ragged2e} 
\newcommand{\x}[1]{{\color{black}#1}}
\newcommand{\y}[1]{{\color{black}#1}}

\title{{\huge De-Idealizing De-Idealization: \\ Beyond Full Reversal}}
\author{{\LARGE Yichen Luo and Eugene Y. S. Chua}}

\date{\normalsize Accepted at \textit{The British Journal for the Philosophy of Science}. \\ Preprint of 25 February 2026. Please cite published version when available.}

\begin{document}

\maketitle
\noindent\rule{\textwidth}{1pt}

\vspace{5mm}

\begin{abstract}
\justifying \noindent There is a question of whether de-idealization is needed for justified use of -- for `checking' -- idealizations. We argue that the standard philosophical account of de-idealization has become too idealized, but that this does not preclude the possibility of justificatory practices which show how models can be used to make inferences about the world. In turn, motivated by examples in physics, we provide a more expansive and practice-driven account of de-idealization by relaxing the standards for closeness to more realistic theoretical items, identifying at least three kinds of procedures for de-idealization: intra-model, inter-model, and measurement de-idealizations. These examples highlight how idealizations can be -- and indeed have been -- scrutinized within physics without appealing to the philosopher's idealized notion of de-idealization.
\end{abstract}

\begingroup
\tableofcontents
\endgroup

\noindent\rule{\textwidth}{1pt}

\newpage 
  \section{Introduction}

In science, idealization introduces distortions into our models of certain systems in the world (what we call `target systems'). The goals of idealization are varied: for example, extracting salient features from fine-grained and complex systems, isolating relevant causal patterns, discarding irrelevant features, or providing computationally simple solutions to problems.\footnote{These disparate functions rest on disparate understandings of idealization. See, respectively, \cite{Craver2006-CRAWMM}; \cite{Potochnik2017-POTIAT-3} and \cite{Pincock2022-PINCSM}; \cite{Strevens2008-STRDAA}; McMullin's Galilean idealization \parencite{McMullin1985-MCMGI-2} and Smith's Newtonian idealization \parencite{smith_closing_2014}.} The domain of physics -- our subject matter of interest in this paper -- provides a treasure trove of such examples of idealizations. In Newtonian physics, we can take the Earth to be a point particle concerning its motion around the Sun; in thermodynamics, the ideal gas is assumed to consist of randomly moving particles that are not subject to intermolecular interactions; in Bohr's semiclassical atomic model, electrons move in a definite orbit around a fixed-position nucleus; in black hole physics, exact metrics are solved by assuming black holes to have certain symmetries and be surrounded by external vacuum spacetime. The list goes on.

Granting that idealization is \textit{rampant} in scientific practice, we are confronted with a pressing epistemological question: why are we authorized to employ these idealizations to make inferences about real systems? \y{After all, idealized models are distorted representations of real systems of interest: they deliberately depart from a fully comprehensive and accurate representation in various respects}.\footnote{\y{``Distortion" here does not assume a naïve correspondence theory of truth, nor that idealizations are literally false, only that they're non-veridical or non-literal in some respects.}} Furthermore, we know many cases where appeals to idealizations failed, such as Einstein's static universe model or the mechanical structure of the luminiferous aether.

Since the myriad idealizations of physics include \y{distortions}, and not all idealizations work, it behooves us to \textit{check} whether inferences about the world made with these idealizations may nonetheless be trusted and relied upon. This process of checking idealizations -- of providing \textit{justifications} for their use in modelling the world -- is what we take to be the heart of \textit{de-idealization}. 

However, the standard construal of de-idealization, often attributed to \cite{McMullin1985-MCMGI-2}, has come under fire from opponents who contend that this strategy may be too demanding, resting on examples from physics and beyond (\cite{Knuuttila2019-KNUDNE}). For instance, \textcite{Rice2021-RICLDE-5} provides a systematic account of idealization which focuses on the positive and ineliminable contributions of idealizations in scientific practice, including explanation, which cannot be de-idealized. \textcite{Bokulich2011-BOKHSM} analyses the semiclassical Bohr model in explaining the hydrogen atom, with which she argues that the Bohr model is a strict idealization -- what she calls a fictionalization -- which nonetheless offers a genuine explanation of the phenomenon. Furthermore, she contends that the Bohr model cannot be de-idealized in terms of e.g. quantum mechanics, yet nevertheless provides a good explanation. \textcite{Potochnik2017-POTIAT-3} offers a more pragmatist argument for understanding the essential roles of idealization in terms of the goals of science (and scientists): idealizations are permanent and ineliminable because little effort from scientists has been put into de-idealizing idealizations, as she says, ``...at least some idealizations are permanent, in the sense that reducing or removing the idealizations to increase the accuracy of representations is not taken to be a goal.'' \parencite[58]{Potochnik2017-POTIAT-3} Although we do not accept the above views wholesale, we agree that they highlight a crucial worry: we may not be able to successfully remove \textit{all} idealizations in every case, yet we must carry on and idealize nonetheless.

We have reasons to take such worries seriously. If de-idealization is too demanding or misaligned with scientific practice, then we agree that it should not be a methodological demand. However, we argue that the contemporary problematization of de-idealization \textit{itself} hinges on a strong, idealized, philosopher's construct which is unnecessary and impractical for justifying scientific practices. 

Our main goal here is to de-idealize the philosopher's notion of de-idealization. We first present the standard reading of de-idealization (\S2). Then, in \S3, we propose a general schema for de-idealization in terms of a model's \textit{closeness} to \textit{more realistic} theoretical items. In the \textit{idealized limit}, this more realistic item is the \y{full representation of some target system}, while closeness can be understood in the usual mathematically rigorous sense of approximation and asymptotic reasoning. However, we propose that both the requirement of a full representation and the characterization of closeness in the mathematically rigorous sense can be relaxed, resulting in a \x{more expansive notion of de-idealization}. Drawing from examples in physics, we highlight three ways of \x{de-idealizing in practice}: intra-model (\S4), inter-model de-idealization within the same domain (\S5.1) and across distinct domains (\S5.2), and measurement de-idealizations (\S6). Through these ways of de-idealizing, we show how de-idealization as a general notion can jettison its idealized baggage, pushing back against criticisms that this notion is not in line with scientific practice or that we need not check our idealizations. At least in physics, and perhaps beyond, de-idealization matters. In turn, we hope to present a different, messier, but no less epistemic notion of de-idealization that is situated in practice. We can find many justificatory strategies that show how idealized models \y{provide reliable inferences about the world}, as long as we are willing to relax what we mean by closeness, and what we mean by a theoretical item being more realistic.

 \section{What is de-idealization?}

The concept of de-idealization is often attributed to \cite{McMullin1985-MCMGI-2} and his `Galilean idealizations'. These obtain when one introduces simplifying or omitting assumptions in one's modelling process (even as one knows or suspects that they are in fact not true), or ignores certain material features of a system, in order to make a model computationally tractable. 

Consider McMullin's example of the Bohr model. \x{In Bohr's 1913 treatment the nucleus is taken to be at rest (or equivalently, infinitely massive), the electron is described as a localized particle moving in stationary elliptical orbits around the nucleus, it is assumed not to radiate while in such orbits, and its motion is governed by non-relativistic dynamics.\footnote{One idealization in the Bohr model discussed by \textcite[260]{McMullin1985-MCMGI-2} is the assumption of circular orbits. However, as an anonymous reviewer helpfully pointed out, while it is true that Bohr assumes circular orbits as a convenient special case of the more general elliptical orbits for one-electron systems, his model did start with an allowance for elliptical orbits, and he uses circular orbits with an explicit note that this ``will, however, make no alteration in the calculations for systems containing only a single electron'' \parencite[4]{bohr_constitution_1913}. Furthermore, just a few years later, Sommerfeld's extension already treats the general elliptical case and confirms that circularity is largely inessential here (cf. \textcite{sommerfeld_quantentheorie1_1916}).} \y{These assumptions were recognized, even when the model was first introduced, to be highly idealized and not strictly accurate in a literal sense.} After all, nothing is infinitely massive, a bound electron's motion need not be circular and electrons in low-lying orbits move at speeds where relativistic effects are expected. Furthermore, according to classical electrodynamics, an accelerating charge should radiate. Nevertheless, these assumptions facilitated calculations and generated predictions. Notably, with these assumptions, and the then-estimated values of the electron mass and charge, the speed of light, and Planck's constant, one can use the Bohr model to predict the Rydberg constant. Since there were already experimental measurements of the Rydberg constant, one could then compare the theoretical predictions to measurements. Crucially, one main reason for the scientific community's initial adoption of the Bohr model was, in fact, how closely the model's predictions approximated these experimental measurements (more in \S6).}

However, the model is clearly \y{distorted} as a whole. How then are we justified in using its predictions to model real hydrogen atoms? McMullin's goal is to push back against \cite{Cartwright1983-CARHTL}'s idea that explanation excludes or has little to do with truth, by showing how scientists can justifiably use idealizations to represent the world even if they contain false assumptions. Enter de-idealization: the means through which one \textit{justifies the use of such idealized models}. For McMullin, de-idealizations are broadly understood as procedures through which simplifying assumptions can be removed or eliminated in a targeted fashion, or where  ignored material features are added back. This is in line with Earman's principle for interpreting idealized models: ``no effect can be counted as a genuine effect if it disappears when the idealizations are removed.'' \parencite[191]{Earman2004-EARCPA} 

\x{One example of de-idealization in this case concerns the Bohr model's non-relativistic treatment of the electron. Bohr's original model treats the electron's motion using Newtonian mechanics and thus ignores relativistic corrections, even though the electron velocities in low-lying orbits are a non-negligible fraction of the speed of light. Sommerfeld's subsequent extensions (e.g. \textcite{sommerfeld_quantentheorie1_1916}) of Bohr's theory de-idealize this assumption by extending the theory to consider relativistic elliptical orbits (among other refinements). The resulting theory refined the energy levels predicted by Bohr and predicted the observed fine structure of the hydrogen spectrum, while reducing to Bohr's more idealized non-relativistic model in the appropriate limit (for $v \ll c$). In this way, we show that results based on the idealized Bohr model didn't essentially depend on the idealized assumption of non-relativistic motion, and that more realistic relativistic models approximate the idealized Bohr model. This, in turn, provides justification for our use of the Bohr model, i.e., is a de-idealization of the Bohr model.}

In the recent literature on idealization, however, McMullin's idea of de-idealization has been met with scepticism. Prima facie, de-idealization appears to be a strong requirement, since we introduced these simplifying assumptions precisely so that one can begin calculating or theorizing about something previously too complex. So the demand that one adds back these complexities is a practically challenging one. For instance, \cite{Bokulich2011-BOKHSM} briefly entertains the worry that ``it is not clear that all idealizations can be subject to a de-idealization analysis", citing the work of \cite{Batterman2004-BATCPA} on the indispensability of infinite idealizations.\footnote{We think Batterman's worry has been largely mitigated by recent work from e.g. \cite{Wu2021-WUEUI, Butterfield2011-BUTLID, Norton2012-NORAAI, Menon2013-MENCPQ}.} 

This practical challenge gets raised to a full-blown problem that de-idealization is \textit{not in line with scientific practice} by \textcite[658]{Knuuttila2019-KNUDNE}. For them, the demand of de-idealization amounts to ``an insidious presumption by philosophers of science that scientists originally start from considering the real world and then arrive at their models through idealizing (and abstracting)", which leads philosophers to think that scientists can easily ``reverse their modelling recipes to get back down to the more fully blown situation they started with, fraught though that process might be." \parencite[659]{Knuuttila2019-KNUDNE} This is especially so since scientists do not start with the ``real world" (presumably with \textit{all} its details) and then proceed to simplify. Adding back details is not as simple as `reversing' these simplifications. Rather, they often start off with simplified models, from which details are added in later. Others like Weisberg also interpret de-idealization as a demand for ``de-idealization back to the \textit{full representation}" (\textcite[646]{Weisberg2007-WEITKO-2}, emphasis ours). The criticism, then, seems to be that a search for de-idealization is completely misguided and out of line with scientific practice. On Weisberg's interpretation of de-idealization, an idealization is only justified if we can return to the full representation by ``removing any distortion, and adding detail back to the model": ``The ultimate goal of Galilean idealization [in this strong sense] is \textit{complete representation}" (\textcite[655]{Weisberg2007-WEITKO-2}, emphasis ours). If what justifies the use of idealizations is \textit{this} strong, `full' notion of de-idealization, then it is unclear whether we ever have such a justification, since it is unclear at any point in time whether we can really obtain a ``full representation'' of the sort demanded. For instance, the Bohr model would \textit{not} be justified back in the 1920s, since we have yet to de-idealize the proton in this full sense. 

Indeed, justification in terms of de-idealization in this sense becomes elusive until we have a `final theory of everything'. To echo recent discussions of effective field theories (e.g., \cite{Williams2019-WILSRM}, \cite{Wallace2022-WALTQT}) and naturalized metaphysics (e.g., \citeauthor{McKenzieforthcoming} (forthcoming)), we simply do not have a fundamental theory with which we can provide such truths; all we have are effective theories. These are theories known to break down at some point -- \y{they involve idealizations and approximations that we expect to fail outside their domains of applicability} -- but \y{without accompanying descriptions of what lies beyond that point.} If de-idealization requires that idealized models approximate the \textit{full representation}, we concede that it cannot be achieved because we never have a grasp on that full representation. 

A distinct but similar argument in the vicinity is given by \cite{Potochnik2017-POTIAT-3}, who appears to attack the very idea that idealizations need to be `checked'. In her own words, idealizations are (and, implicitly, should be) both \textit{rampant} and \textit{unchecked}. We have no qualms with how rampant idealizations are, so let us focus on her argument that they are (and, implicitly, should be) unchecked, in the sense that \textit{they do not require de-idealization} at all. Here it must be emphasized that her target is, again, the `full' sense of de-idealization. Potochnik essentially argues from scientific practice: there is in fact little focus (by scientists) ``on eliminating idealizations or even on controlling their influence." \parencite[58]{Potochnik2017-POTIAT-3} Her argument comes in two parts. Firstly, she argues against eliminating idealizations: ``at least \textit{some} idealizations are permanent, in the sense that reducing or removing the idealizations to increase the accuracy of representations is not taken to be a goal", especially when these idealizations best ``facilitate the representation of core causal influences" or when we cannot eliminate the assumptions due to absolute computational or cognitive limits \parencite[58]{Potochnik2017-POTIAT-3}. Secondly, she argues that there is furthermore \textit{little emphasis in practice} for controlling the influence of idealized assumptions. However, upon closer inspection, what she really claims is that there is little emphasis in practice for controlling the influence of idealizations \textit{in a very specific sense}. This sense of de-idealization would require a full demonstration that ``idealized factors... be unimportant to render idealizations unproblematic, that is, to keep them in check." \parencite[58]{Potochnik2017-POTIAT-3} This is the heart of Potochnik's contention that idealizations are `unchecked': she contends that scientists do not de-idealize idealizations by showing that they are \textit{justified in leaving out all causally irrelevant features}, i.e. by showing that \textit{all} features left out are \textit{in fact} unimportant. This amounts, again, to a demand that we return idealizations back to the ``full representation'', something that Potochnik contends. For her, scientists simply \textit{do} leave out these features, rather than `checking' whether they are unimportant. 

 \section{De-idealizing de-idealization}

The foregoing criticisms are well-taken, but are they actually problems for de-idealization? We think that these criticisms are correct in calling out a certain idealized notion of de-idealization, which we call \textit{full de-idealization}, to echo Weisberg: this is de-idealization which aims to justify an idealized model by providing ``full representations" of the target system. Only in \textit{this} sense, we agree that idealizations do not need to be de-idealized. In our view, however, we agree with \textcite{Knuuttila2019-KNUDNE} that full de-idealization is a philosopher's construct, for a simple aforementioned reason: we \textit{never} have the full representation of the world, and all we have are models. Yet, to reject full de-idealization does not amount to rejecting the broader epistemological project of de-idealization: a search for justification for why our models can be used to make reliable inferences about the world. 

\y{Critics of de-idealization have often turned instead to pragmatic considerations for accepting or rejecting models.\footnote{See, e.g., \cite{Weisberg2007-WEITKO-2, Bokulich2011-BOKHSM, Potochnik2017-POTIAT-3, Knuuttila2019-KNUDNE}.} While we see the value of such analyses, we think it is too quick to give up checking whether, and how, our models are constrained by the world, just because one specific relation -- full de-idealization -- fails to obtain.\footnote{This is not to deny \textit{all} pragmatic elements. As we discuss below, modelling goals play a central role in determining what counts as a successful justification (see, e.g., \cite{Giere2009-GIEAAC-4}).} Scientific models must remain answerable to the world in some substantive sense: some function as free-standing representational devices, while others support reliable inferences about the world. De-idealization, on our view, is best understood not as a demand for literal model-to-world correspondence, but as strategies for justifying that a model's successes are not accidental -- that its idealizations are appropriately constrained by, and responsive to, features of the target system rather than unconstrained artifacts of representation.}

The search for a less idealized account of de-idealization need not take us further than McMullin's very own discussion. Far from demanding a return to the full representation, or ``starting with the real world", we can read McMullin's proposal in an attenuated way. On this weaker reading, justifying the use of an idealized model does \textit{not} demand an immediate removal of \textit{all} simplifying assumptions, all at once, or a return to some mythical `full' representation of the `real world'. As McMullin notes himself, ``the model can be improved by \textit{gradually} adding back the complexities." (\textcite[261]{McMullin1985-MCMGI-2}, emphasis ours) 

As McMullin's discussion of the Bohr model exemplifies, the process of de-idealization can be slow, piecemeal, and, importantly, \textit{might involve other idealizations yet}! Later developments of the Bohr model added in more realistic considerations such as relativistic effects, which led to, among other things, Sommerfeld's discovery of the fine-structure constant. These are de-idealizations (what we'll later call intra-model de-idealization), but nevertheless did not \textit{wholly} de-idealize the Bohr model: the proton was still assumed structureless. What is \textit{inside} the proton wasn't even known or theorized about until much later, with the advent of quantum chromodynamics. 

In other words, weaker de-idealizations can be performed by relying on a \textit{less} idealized, not \textit{un}-idealized, model. Aspiring de-idealizers need not -- and indeed never -- ``start from the real world" or the ``full representation"; they work from \textit{within} models and try to check them, by developing \textit{other} models. To de-idealize an idealized model, we can remove some -- but not all -- idealizing factors in a partial, piecemeal way, without requiring the removal of all idealizing factors: de-idealization can be effectively justified \textit{only within some assumed regime of validity}, not tout court. Indeed, we see the task of de-idealization, in part, to be the task of clarifying the boundaries of a model's applicability, within which it is justified. On such a view, idealized models can be justified to the extent that we can de-idealize -- in the limit they are completely reversed to the ``full representation'', but there's no expectation of \textit{that} sort of story before we can begin taking them to be justified to some extent, by understanding the extent to which an idealization works. This means circumscribing, more or less, the appropriate domains in which they do work. 

Instead, we suggest an attenuated reading of McMullin's view: \x{(a) the idealized Bohr model acquired some justification in the 1920s for modelling the hydrogen atom, because (b) the model could be shown, in one specific mathematically rigorous way, to be \textit{close to} (c) a specific set of \textit{more realistic} theoretical items, a family of models parametrized by e.g. how close the electron's motion is to the speed of light. (More on the Bohr model and why it ended up being abandoned nonetheless, in \S6.)}

Read as such, ``closeness" and ``more realistic" remain terms of art. In the idealized limit of de-idealization, the sort often discussed by its critics, ``closeness" and ``more realistic" are cashed out in terms of rigorous asymptotic reasoning plus robust metaphysics respectively: the use of precise mathematical limits which demonstrate with certainty that an idealized model converges to the ``full representation" of the world itself. In this idealized limit, we have full de-idealization: a precise, robust, sense in which the idealized model closes in on the world itself. But we have already suggested that, a full representation is out of reach in practice; furthermore, as we discuss in \S4, the mathematical procedure of asymptotic reasoning, though powerful, is not always available. Idealized full de-idealization is thus rarely in line with scientific practice.

Nonetheless, this does not mean de-idealization as a notion -- and its role as justifier of idealization -- should be abandoned altogether. Rather, we suggest that there are distinct, weaker, ways to understand the twin ingredients of "closeness" and ``more realistic", without appealing to asymptotic reasoning or the full representation respectively; correspondingly, there are distinct, weaker, ways to understand the de-idealization of idealized models which can show how they are close to things we take to be more realistic. If we are willing to relax our standards for what counts as close or more realistic -- as we think we should, and as we think physicists have done -- then we can see a rich array of justificatory practices in play. 

Four points of clarification. First, an idealized model can be shown to come close to different sorts of \textit{more realistic} theoretical items; for generality we will call them \textit{de-idealizing constructs}. What matters is that the de-idealizing construct is taken to be a description of the target system in relevant ways. We already discussed the de-idealization of an idealized model vis-a-vis other models of a target system \textit{in the same parameter-family}. However, we will also discuss how idealized models can come close to other families of models \textit{not related via sets of parameters}. Furthermore, we will also see that idealized models can come close to not other models, but \textit{measurement outcomes} -- nowadays recognized, rightfully, to be theoretical items in their own right \parencite{sep-measurement-science}.\footnote{Some like \textcite{vanFraassen2008-VANSRP-2} might think of measurements as ``surface models" as well, in which case one can safely think of de-idealization as generally providing a model-to-model relation instead.} By focusing on model-to-theoretical-item relations, this approach lets us avoid being mired in questions about how idealized models represent the world itself, so as to study the more tractable question of how idealized models close in on more realistic theoretical items.

Second, when it comes to the question of what it means for something to be ``realistic", we emphasize that more realistic de-idealizing constructs need not be \textit{wholly} realistic -- they can be idealized in their own ways as well. What matters is that we have prior theoretical reasons -- which will depend on the de-idealizing construct in question -- to believe that these constructs are \textit{more} realistic than the idealized model, in describing some target system. Since this judgement must depend on the particular assumption in question, any concrete standard for being ``more realistic" must depend on the context and purpose of modelling. It is notoriously difficult to spell out what a context is, but it does not prevent us from specifying some common and important features in the context of physics: non-exhaustively, the relevant regimes and scales, which features of the target system we are interested in modelling, and defeasible theoretical reasons for believing that the de-idealizing construct is more realistic -- all play a role in determining what is more realistic. 

Third, a standard for ``closeness" also depends on the context. We take a demonstration that a model is close to a more realistic de-idealizing construct as showing that the former is \textit{close enough} to the latter in the \textit{relevant} respects. What counts as \textit{relevant} depends on the \textit{purpose} for which the model is being employed. Imagine a stick-figure drawing of a person. If your goal is to study basic body motion at a coarse-grained level, such as how the arms and legs swing while walking, this model is close enough in the relevant ways, since it captures joint positions and overall posture. But if your goal is to read fine-grained muscle movements, it fails. Similarly, if your goal is to study facial expressions or detect subtle emotions, the same stick-figure fails entirely. The stick-figure is close enough to a person for studying a person's gait at a coarse-grained level, but not for studying the same gait at a fine-grained level, nor for studying their expression.

What counts as \textit{close enough} is likewise dependent on the context. Consider the idealization $\sin \theta = \theta$, which \y{does not hold for any} $\theta \neq 0$. Whether the associated error term -- which can be ascertained up to precision at arbitrary orders via the Taylor series -- is small enough for us to accept $\sin \theta = \theta$ as a close enough description of a target system must depend on the context of modelling. The error for displacements on the order of arcseconds is on the order of $5 \times10^{-6}$ arcseconds, while the error for displacements on the order of ordinary seconds is on the order of $0.001''$. In astronomy, the former is negligible in all circumstances, while the latter ``does not quite match the accuracy of the most refined astrometric measurements." \parencite{green_spherical_1985} What counts as close enough depends on the goal of modelling, of course. However, note that context-dependence is \textit{not} arbitrariness: the fact that astrometric measurements become noticeably erroneous at the latter scales but not the former is a feature of the world and our epistemic practices, not a matter of convenience or taste. 

Finally, it is important to emphasize that we do \textit{not} claim that we \textit{should} use the de-idealizing construct (such as a refined, more realistic model) in practice \textit{over} the idealized model once we obtain this de-idealization: it may be much easier, expedient, or even more explanatory in certain contexts to continue employing the idealized model over the de-idealizing construct. That is, pragmatic factors might suggest the continued use of idealizations, even after we have obtained such de-idealizations. But these pragmatic factors do not override the need to \textit{justify} why we are allowed to use idealized models. We also leave it open that there may be other de-idealization strategies, ripe for the picking by other philosophers of science.

If we can show that an idealized model approximates more realistic de-idealizing constructs, then the model earns a degree of justification -- not because it corresponds, in any strict sense, to the full and complete structure of the world, but because it stands in closeness relations to descriptions that we have reason to think are more realistic in relevant respects. This suggests a schematic model for de-idealization, which can then be filled in depending on what one means by ``closeness". In what follows, we emphasize that there are at least four distinct, non-exclusive, ways that models can be close to more realistic de-idealizing constructs. 

\x{Before moving on, we want to set aside one possible worry about our account of de-idealization. Some might think that our account of de-idealization is too permissive. Here, we prefer to take a more ecumenical stance and take the heart of de-idealization to be establishing an idealized model's closeness to more realistic theoretical items. This approach is what helped us identify the various procedures later -- a sign of the fruitfulness of such a stance. That said, we can see the motivation behind a more restrictive notion of de-idealization as well. For instance, one might think that de-idealization necessarily requires the refinement of an idealized model in terms of \textit{another model} with the \textit{same target system}.\footnote{We thank an anonymous referee for pointing this out.} Given such a restricted view, we are happy to grant that only the first two procedures we discuss later are proper de-idealizations, while the latter two are distinct justificatory procedures for validating the use of idealized models, and need not be called de-idealizations. Nonetheless, we think they are all still -- and this is what really matters to us -- epistemic practices for validating and justifying the use of idealized models; we thus believe our paper can be useful even to those with more restrictive views of de-idealization.}

\section{Intra-model de-idealization}

The first approach to de-idealization is what we call intra-model de-idealization, because we are supposed to constrain our attention to \textit{within} a single family of models parametrized by some parameter. Intra-model de-idealization demonstrates closeness by employing a mathematically rigorous, notion of approximation via \textit{asymptotic reasoning}.\footnote{See \textcite{Batterman2004-BATCPA, Bender_Orszag_1999}.} Asymptotic reasoning provides techniques for investigating limiting behaviours: instead of solving the full equations of the system, it appeals to the limit value (e.g. 0 or $\infty$) of a certain parameter in the equations, around whose vicinity the local behaviour of the system become simplified and hence tractable. Put simply, standard asymptotic methods turn a complicated function $g(x)$ -- which may be a more complete description of some target system -- into a simplified function $f(x)$ -- an idealized model -- such that in some limiting regimes ($x\to x_0$), the difference between them is mathematically negligible as $\lim_{x\to x_0}\frac{f(x)}{g(x)}=1$. In this way, we see sharply that models with more realistic -- finite -- parameter choices (for $\infty$ is surely not realistic) mathematically converge to a model with an idealized parameter choice. In turn, we can see how the idealized model is de-idealized in terms of models for which $x \neq x_0$. Since the idealized model approximates any more realistic model with error terms up to arbitrary precision, it is clear that the idealized model is close enough to the more realistic model up to any arbitrary standard of closeness. 

Intra-model de-idealization is not novel, but our point is simply that \textit{even} intra-model de-idealization -- the gold standard for McMullin-style de-idealizations -- does not require ``full representations" to succeed; only that we can demonstrate closeness to more realistic models. A classic example of intra-model de-idealization in physics is the (extensively discussed) ideal gas model. In such a case, the target system is some real gas in a box, say, nitrogen or oxygen. We want to describe this gas in terms of an idealized model: the ideal gas equation of state $PV = nRT$. The ideal gas model assumes several idealizations: gas particles do not interact with each other (no intermolecular forces), and gas particles occupy no volume. Since we know that these are false assumptions from a background theory of matter, strictly speaking there are no gases which can be described by the ideal gas equation of state. 

Yet we \textit{have} found out when this model works well, nonetheless, by identifying hidden idealized parameter choices in the ideal gas equation of state. Indeed, the unearthing of these parameters is what led us to realize that the above idealizations hold, to begin with. Most famously, the van der Waals equation of state highlights two hidden parameters $a$ and $b$ in the Ideal Gas Law to be manipulated, which can account for the volume of gas particles and the intermolecular forces, $(P + \frac{a n^2}{V^2}) (V - nb) = nRT$. Here, $a$ and $b$ are constants specific to the gas, representing the strength of intermolecular forces and the finite volume of the gas particles, respectively. In this case, intra-model de-idealization provides justification for the Ideal Gas Law in the following way. First, it explicates how some parameter choices in the ideal gas law were implicitly idealized \textit{with respect to} a broader, more realistic (nonzero), set of parameters $a, b$. Then, intra-model de-idealization specifies how models with more realistic parameter choices converge to that of the model with an idealized parameter choice 
(here, as $a \to 0, b \to 0$, picking out regimes of relatively low pressure and high temperature), rendering mathematically precise the conditions that must hold for the idealization to be approximately suitable for modelling realistic systems. Grasping these conditions allows one to gain new reasons for justifiably applying the ideal gas model: precisely in low-pressure high-temperature regimes.

It is true that the van der Waals equation of state is still idealized, and we need to de-idealize it as well to find out when it works and when it does not -- e.g., it still assumes particles are featureless hard spheres. For that task, we need more sophisticated equations of state or statistical-mechanical models. Nonetheless, the van der Waals family of models is uncontroversially a more realistic model for real gases. First, it is \textit{qualitatively} accurate in certain regimes, in predicting liquid-vapor phase transitions for instance \parencite[10]{epstein_textbook_1937}. That is, it (qualitatively) predicts, describes, and explains the behaviour that is experimentally observed, which the ideal gas law does not. Secondly, despite its significant and well-known quantitative inaccuracies, it is still \textit{more} quantitatively accurate than the ideal gas law, in predicting e.g. pressure-temperature curves of actual gases. (The latter point is due to what we will call, in \S6, measurement de-idealization.)

While the case of de-idealizing the ideal gas is well-discussed, we emphasize that the range of cases classified as intra-model de-idealizations extends beyond it. Another paradigmatic case is justifying the use of the thermodynamic limit in critical phase transitions (e.g., \cite{Norton2012-NORAAI}, \cite{Wu2021-WUEUI}, and \cite{Lavis2021-LAVBLB}), where a deeper understanding of renormalization group methods shows how it is unnecessary to assume an infinite number of molecules, implying the dispensability of the thermodynamic limit; a large but finite number suffices to account for the phenomena.

However, importantly, even though asymptotic reasoning offers the precise and rigorous notion of closeness, it is not always available. As many in the literature have pointed out, we do \textit{not} always possess such a clear and precise mathematical grasp on our idealized models. This problem is especially prominent in modelling practices at the frontier of physics, particularly in research on gravitational phenomena or within the quantum realm. Models in these domains often rely on various idealizations, such as spacetime symmetries, which \textit{cannot} be intra-model de-idealized either for practical reasons related to computational feasibility or for fundamental reasons concerning mapping idealizations into parameters. 

First, we might not be able to target and intra-model de-idealize any particular modelling assumption because the assumptions are made in a complicated, interconnected, way (echoing \textcite{Rice2018-RICIMH} about ``holistic distortions" -- assumptions which cannot be targeted individually and de-idealized like $a, b$ in the ideal gas law). Second, even if we \textit{do} find such a parameter with which to relate idealized models to more realistic models, we might still fail to obtain the sort of convergence required by asymptotic reasoning, if variations of the parameter leads to `blow-ups' or singularities. For instance, we can try to vary the cosmological constant $\Lambda \to 0$ in order to show how flat spacetime approximates de Sitter (constantly, positively, curved) spacetime and to de-idealize flat spacetime that way. However, this leads to discontinuities in energy \parencite[2--3]{PhysRevLett.116.051101}, so that we cannot say that there is convergence of a sequence of non-flat spacetimes to flat spacetime in this regard. Third, as \textcite{Chua2025-CHUNQK} has emphasized recently, there is no clear way to de-idealize the Killing symmetries assumed in various idealized models of black hole physics, in this strict sense of intra-model idealization (cf. \citeauthor{RyderForthcoming-RYDTBH} (forthcoming)). Finally, even when we do find intra-model de-idealizations, they are often provided long past the point where we would have needed some initial justification for their use in modelling.

\section{Inter-model de-idealization: within a domain, across domains}

\subsection{Inter-model de-idealization within the same domain}

An immediate question arises: If a model cannot be intra-model de-idealized and is not related by parameter variation or limiting procedures to more realistic models within the same parameter-family, can its use still be justified? In such cases, we suggest that we simply need to relax our notion of closeness between an idealized model and more realistic de-idealizing constructs. Features of the (family of) idealized models can be shown to be close to features of some other more realistic family of models, \textit{even if} we cannot prove the sort of asymptotic convergence demanded by intra-model de-idealization. In what we call \textit{inter-model de-idealization}, we do not target specific assumptions and then de-idealize, for reasons already mentioned. Instead, we justify the use of an idealized model to model certain \textit{features} of the target system, by establishing that the features of an idealized model is close enough to a more realistic family of models by identifying \textit{conceptual continuity} between idealized and more realistic families of models. By conceptual continuity, we mean that the target feature in the idealized model can be found, more or less, in the more realistic family of models (which serve as the de-idealizing constructs in this case). 

To explicate inter-model de-idealization within the same domain, and the notion of conceptual continuity, let us consider the idealized models of classical black holes (as exact solutions). The epistemological question here is whether we can justify the use of idealized black hole models, such as the Schwarzschild model, in modelling target systems with, say, the features of an event horizon and singularities. Here we briefly follow the case study done by \textcite[Ch.2]{Luo2025}. 

In modelling classical black holes, there are at least three approaches which utilize different mathematical techniques to explore the physics of black holes within the framework of general relativity (GR): the exact solution approach, the topological-causal approach, and the initial-value approach. Firstly, the earliest models of general relativity were exact solutions, i.e., explicit, analytic solutions to Einstein's field equations derived under specific idealizations such as spherical symmetry or asymptotic flatness. Secondly, the topological-causal approach introduces topological methods to black hole physics, enabling more general characterizations of global causal structures of Lorentzian manifolds, establishing a hierarchy of causal conditions in spacetime, and identifying shared structural properties across black hole solutions. Finally, the initial-value approach reformulates GR by examining Einstein's field equations from the perspective of Cauchy problems, using techniques from partial differential equations (PDEs) to specify a complete set of initial data (as dynamical variables) governing the causal evolution of solutions, facilitating the study of black hole dynamics.

In the exact solution approach, black holes are represented as stationary states of gravitational collapses, by the family of Kerr-Newman models, such as the spherically symmetric Schwarzschild model. Strong idealizations are assumed in such an approach to modelling: to get exact solutions of the non-linear field equations, black hole solutions are required to be stationary, and to satisfy various symmetries, such as spherical symmetry or axial symmetry. Crucially, such idealized models entailed two features of black holes: \textit{event horizons} which mark the boundaries of no return, and \textit{singularities} which signify the infinite gravitational strength inside black holes.  

For a long time, the two properties were thought to be artifacts from imposing these idealizations, rather than features of real-world target systems. Furthermore, within the exact solutions approach, the above idealizing assumptions can hardly be mapped into controllable parameters. Even if the mapping \textit{were} possible, intra-model de-idealization procedures are extremely limited by the non-linearity of the field equations. Indeed, before development of the topological-causal approach, this doubt was significant in undermining justification for believing in the existence of black holes.

Though not intra-model de-idealization, the topological-causal approach provided new resources for de-idealization by bypassing the need to directly solve the field equations, instead offering a systematic analysis of the causal structures of black hole spacetimes under generic conditions shared across the space of black hole solutions. For instance, rather than understand singularities in terms of divergent curvature, we understand them via mathematically \textit{distinct} objects -- inextendible causal geodesics of finite affine length.\footnote{Of course, again, we can debate whether inextendible causal geodesics are close ``more or less" enough to curvature blow-ups, but that will take us too far afield. We stick with the standard physicist's stance here, but see \parencite{Earman1995-EARBCW-3} for the locus classicus on the matter.} Notably, this approach enabled the Penrose-Hawking singularity theorems (e.g., \cite{Penrose1965}; \textcite[Ch.8]{HawkingEllis1973}), demonstrating that the spacetime singularities of black hole solutions are not mere artifacts due to symmetry assumptions in the exact solution approach.\footnote{Concerning event horizons, for example, \cite{Finkelstein1958} showed that pathological features related to event horizons were merely coordinate-singularities which can be overcome by a maximal extension of the spacetime. Furthermore, in light of the singularity theorems, the globally defined event horizons can be replaced in many situations by more suitable, locally defined notions such as trapped surfaces.} Rather, they are generic to classical black hole spacetimes in GR. This is especially because the singularity theorems are proven without appeal to the symmetries assumed; they only impose generic conditions on spacetimes, such as a non-negative energy density (null energy condition), the absence of closed time-like curves (causality condition), and the existence of \textit{closed trapped surfaces}, which signify the focusing effects characteristic of black holes. 

Here we see conceptual continuity of the notion of ``singularity." In the idealized model, singularities are represented by curvature blow-ups. In topological-causal models, it is instead represented by inextendible geodesics of finite length. Nonetheless, the two notions of singularities are conceptually continuous, even though there does not exist a necessary connection between them in every physical situation (\cite{Weatherall2023}). Both involve a breakdown in the predictability or completeness of the spacetime manifold. Both can be used to indicate gravitational collapse, and both signal a fundamental limit to classical general relativity. Additionally, each implies a kind of boundary beyond which classical descriptions fail: for curvature blow-ups, this is where curvature invariants diverge; for inextendible geodesics, this is where causal paths simply cannot be continued. Furthermore, mathematically they make similar predictions, in the asymptotic regimes where the idealized model can be taken to hold (such as in the Schwarzschild spacetime). 

Granted, they differ in formulation and emphasis. Curvature blow-ups are metric-dependent and hinge on the behaviour of curvature tensors/scalars, while inextendible geodesics are coordinate-free and rely on causal and topological structures. Moreover, curvature singularities require a notion of divergence, often involving limiting processes, whereas geodesic incompleteness is a more global condition. Whether these differences can be neglected depends on the specific physical scenario of interest: if, in some contexts, a version of the singularity theorem applies and a divergent curvature also appears, we then possess different and compatible means to check the \textit{same} feature. It is in this sense that we take the conceptual continuity of the `singularity' concept across models to demonstrate the closeness, more or less, of the original concept (curvature blow-ups) in the idealized model to the new concept (inextendible geodesics) in the more realistic model. A similar story can also be given for the conceptual continuity of the `horizon' concept, from the event horizon in the idealized model to closed trapped surfaces in the topological-causal approach.

However, proving the singularity theorems crucially requires the \textit{existence} of closed trapped surfaces. Yet, due to their mathematical abstractness, there was no evidence for whether closed trapped surfaces could exist in reality, and if so, whether they could form (or ``evolve") in more realistic cases. That is to say, even though we have de-idealized the exact solutions of general relativity in terms of models of the topological-causal approach, we can also raise the same question of de-idealization for the latter \textit{again}. (But note that the question has shifted! It is no longer about justifiably using symmetries despite their not being realistic; rather, it is now about whether we can justifiably use the singularity theorems despite its reliance on the existence of \textit{trapped surfaces}).

Problems like these motivated the study of black hole dynamics in the initial-value approach. In this approach, it is finally shown, due to the development of advanced PDE techniques \parencite{Christodoulou2009}, that initial data sufficiently give rise to the \textit{formation and evolution} of closed trapped surfaces \textit{under generic conditions}. That is, under generic conditions (crucially, conditions which do \textit{not} undermine the original topological-causal approach's recovery of horizons and singularities), using generic initial data under the general relativistic dynamics, we can motivate the existence of trapped surfaces, with which the singularity theorems follow, rather than assuming them as in the topological-causal approach. This, in turn, de-idealizes the topological-causal approach -- while adding even \textit{more} justificatory strength to the original goal of using the horizons and singularities of the exact models. 

By providing distinct techniques -- analytic solutions, topological-causal reasoning, and numerical relativity -- we can explicitly see conceptual continuities from the notions of event horizons and singularities, to the existence of closed trapped surfaces, and to the formation and evolution of trapped surfaces. Focusing on this conceptual continuity generates a deeper understanding of classical black holes as stable and robust entities whose horizons and singularities do not depend essentially on idealized assumptions. Not only that, grasping such conceptual continuity also gives rise to an \textit{integration} of the three approaches: for example, absorbing physical insights from the exact solution approach, causal structures of black hole spacetimes, and PDE techniques, numerical relativity nowadays can even study black holes in complicated environments (such as black hole mergers) which give rise to observational evidence of black holes (e.g. the LIGO detection of gravitational waves from black hole mergers).

We believe inter-model de-idealization can justify the use of an idealized model in representing a target system, in two senses. First, the continuity of features or concepts across the idealized model and other models shows the \textit{stability} of those features of the idealized model:\footnote{This is deeply connected to the vast literature on robustness, which we won't get into here. See e.g. \cite{Winsberg2021-WINWDR, Stegenga2017-STERAI-8, Wimsatt2012-WIMRRA-2}.} the feature of interest in the idealized model will not vanish, more or less, even under different modelling techniques and assumptions. Second, by tracing conceptual continuity across different models, distinct mathematical techniques are used to examine the associated assumptions, conditions, or properties that contribute to the persistence of the same feature or concept. This allows modelers to have a refined grasp of those features of the original idealized model which are significant for modelling the target system. In inter-model de-idealization, an idealized model earns justification because its core structural or dynamical features can be tracked across increasingly more realistic families of models, even if models are not related by parameter variation or limiting processes as per usual discussions of de-idealization or asymptotic reasoning. The approximation at stake is \textit{not} one of mathematical convergence, but of showing that certain idealized features are preserved, more or less, across a web of more realistic constructs.

The general lesson here is that intra-model de-idealization is a good way, but not the \textit{only} way to demonstrate the closeness of one model to another, and hence for de-idealization. Rather than search for the right parameter to vary, the search for inter-model de-idealization within the same domain instead drives a search, and interplay, of independent modelling strategies for the same target system. The aim here is explicitly \textit{not} to demonstrate mathematical convergence in the sense of asymptotic reasoning; rather it is to unearth conceptual continuity across different, increasingly more realistic, approaches. This, in turn, provides justification -- in appropriate regimes and for appropriate features -- for using the original idealized model as a model of the target system.

\subsection{Inter-model de-idealization across distinct domains}

We may not always have independent modelling strategies in a single domain for the same target system. In that case, we are not privy to both intra-model and inter-model de-idealization in some domain. How then are we supposed to justify the use of an idealized model?

Here we identify yet another strategy that can play this role, which we call \textit{inter-model de-idealization across distinct domains}. Similar to inter-model de-idealization within a domain, the aim is to justify an idealized model by demonstrating how features of an idealized model are conceptually continuous across different families of models, which may not be related by the variations of any particular parameter. Rather, the conceptual continuity is between families of models which may be in \textit{different} domains.

Because they don't have the same domains, the establishment of conceptual continuity is complicated: unlike the prior case where the different techniques are within the same domain and hence share physical meaning (e.g. that the models are all about the very same phenomenon or object and motivated by related physical problems), there is no guarantee that the different families of models are about the same thing at all. Then, the demonstration of conceptual continuity may fail not only because we fail to find mathematical similarities at all, but also because these mathematical similarities -- even if found -- may be \textit{merely} formal.  

Nonetheless, two typical features of inter-domain inter-model de-idealizations can be identified in physics. First, there must be formal similarities between features of the idealized model and features of the de-idealizing construct. Second, these formal similarities must be interpreted and given physical meaning: they must not only be merely mathematical coincidences, but must also be reasonably physically similar in certain ways. Then, we are justified in using the idealized model to model said features, in virtue of how it approximates -- via formal and physical similarities -- features in a different model across different domains.\footnote{One may appeal to accounts of analogical reasoning \parencite{Norton2021-NORTMT, Hesse2000-HESMAA-7, Bartha2010}. However, our aim here is to characterize this construct as broadly as possible, without being tied to any particular account of analogy.} 

There are many physical examples which follow this line of inter-model de-idealization.\footnote{See, e.g., recent work by  \cite{Lehmkuhl2024, gomes_rovelli_2024}), some more controversial than others (e.g. the debate over so-called `analogue gravity', see \cite{Crowther2019-CROWWC}).} To illustrate the de-idealization strategy, we focus on a well-established case discussed by \cite{gomes_rovelli_2024}: the inter-model de-idealization of asymptotic flatness in highly idealized spacetime models of general relativity (GR). In this case, the feature that \textit{gravitational} waves carry energy is modeled by means of a de-idealizing construct drawn from \textit{electromagnetic} wave models. The goal is to justify the use of these idealized models in predicting that gravitational waves carry energy, by establishing that these waves are formally and \textit{physically} similar to electromagnetic waves which do carry energy and are features of a model in classical electromagnetism we \textit{already take to be more realistic for independent reasons} (i.e. de-idealizable in various ways).

Whether gravitational waves carry energy has remained an open question in the literature.\footnote{See \textcite{Duerr2021-sf, HOEFER2000, doi:10.1086/662260}, \parencite[Ch. 7]{Cartwright2022-CARTTO-46}.} On the one hand, gravitational waves formally behave in ways similar to electromagnetic waves,\footnote{For example, they both travel at the speed of light along null directions, possess transversal polarizations and oscillatory behaviours, are radiated by sources with wave-like fall-off rates, etc. Moreover, both the Einstein field equations and Maxwell equations are quasi-linear hyperbolic equations.} the latter being widely accepted to be carriers of energy. And in recent times, we have finally managed to detect gravitational waves, as we did electromagnetic waves. However, on the other hand, scepticism about gravitational energy arises because it can only be precisely described and investigated in highly idealized models. 

To be precise, making sense of gravitational waves as energy carriers requires, first, a consistent definition of energy, and, second, an ability to separate the radiative/wave part from the source/Coulomb part of gravitational fields. Both are highly non-trivial in GR: there is no locally meaningful notion of energy conservation, and a well-defined separation of different parts of gravitational fields cannot be made at finite distances between sources and waves -- only at asymptotic infinity.

To address the above difficulties, spacetime solutions are assumed to be \textit{asymptotically flat}, roughly, that spacetime is exactly flat at infinity. This assumption is uncontroversially an idealization.\footnote{For instance, \cite{Duerr2019-DUEFBA}: ``[...asymptotic flatness is] an idealization in Norton's sense: The embedding spacetime is an unrealistic, surrogate spacetime." See also \textcite[\S6]{Chua2025-CHUNQK} for a discussion of the challenges facing intra-model idealization of asymptotic flatness.} Imposing the idealization of asymptotic flatness allows us to model a dynamically isolated spacetime, which, in turn, allows us to define fruitful global notions of energy such as the ADM or the Bondi mass \parencite[4]{gomes_rovelli_2024}. It also gives spacetimes algebraically special properties and the use of the Newman-Penrose formalism.\footnote{For instance, the property of conformal compactification.} These properties enable the separation of the radiative from the Coulomb parts of the gravitational fields, while also leading to the famous peeling theorem for gravitational radiation: the fall-off rate is the appropriate relation of $\sim \frac{1}{r}$, \textit{exactly the same} as the peeling theorem for the electromagnetic tensor, in the asymptotically distant `wave-zone.' Therefore, the idealization of asymptotic flatness provides an important idealized regime in which the behaviours of gravitational radiation can be meaningfully and unambiguously demonstrated. Furthermore, \textit{within} idealized asymptotically flat models, we can see that there are many significant senses in which gravitational waves formally resemble electromagnetic waves. 

But all these similarities between electromagnetic and gravitational waves are, for now, still formal; it is not yet shown that these mathematical similarities enable us to physically interpret the energy of gravitational waves the same way as the energy of electromagnetic waves. This is especially because the energy of electromagnetic waves is generally covariant, while the energy of gravitational waves is, infamously, \textit{non}-covariant in standard derivations. 

Gomes and Rovelli argue that gravitational waves are not only formally similar to electromagnetic waves in this idealization, but really are like electromagnetic waves as a matter of physical behaviour even beyond this idealization. Their argument can be seen as a combination of the de-idealization strategies discussed in this paper, intra-model de-idealization and measurement de-idealization (discussed later), showing how these strategies need not be independent of each other. Crucially, they argue that the de-idealization strategies that worked for electromagnetic waves can be shown to work for gravitational waves as well. Not only is the idealized gravitational wave model formally similar to the idealized (but more realistic) electromagnetic wave model, but their \textit{de-idealizations are formally similar as well}. This motivates, for them, a stronger similarity than mere formal similarity -- a physical similarity. Caveat: we can of course debate whether this physical similarity \textit{really} holds, but to the extent we can establish physical similarity between an idealized model and a more realistic de-idealizing construct across different domains, with regards to some feature, to that extent we are justified in using the idealized model to model said feature despite the idealizations. 

Gomes and Rovelli's analysis unfolds in two stages. First, they show that in the Newman-Penrose formalism, (components of) the Weyl tensor $\Psi_{4}$, which asymptotically captures the radiative degrees of freedom of the gravitational field, does not in fact essentially depend on flatness \textit{at} infinity. Although $\Psi_{4}$ is precisely defined at null infinity in asymptotically flat models, its value changes only marginally as one retreats to any sufficiently large but finite radius. As they write: ``[For] each particular spacetime model of the scenario, we can compute $\Psi_4$ at ever farther geodesic distances from the source, on a frame that has a well-defined physical interpretation, and verify that its difference to the value taken asymptotically is bounded by some relevant (e.g. experimental) limit." \parencite[8]{gomes_rovelli_2024} This is an intra-model de-idealization: varying the radial position $r$ in $\Psi_4$ for the family of asymptotically flat solutions, we discover that the relevant observable converges as $r \to \infty$.\footnote{Now, crucially, asymptotic flatness as an idealization is still assumed; what matters here is that $\Psi_4$ remains well-defined outside of the asymptotically flat region.} Hence the idealization of flatness at infinity can be relaxed. 

Now, crucially, they point out that this particular intra-model de-idealization is formally analogous to the one used in de-idealizing electromagnetic waves, when we replace the strictly far-field $1/r$ zone by a region that is merely ``far enough" for the Poynting flux to stabilize. Because both theories not only adopt the same idealization that they are only well-defined in the far field, but \textit{also} adopt the same intra-model de-idealization strategy in terms of the \textit{same} formal fall-off hierarchy (i.e. behaviour in the near, transition, and radiation zones), they suggest that gravitational waves are physically similar, not just formally, to electromagnetic waves. 

In the second stage, they suggest that the errors in \textit{empirical} measurements of gravitational waves is controlled by the foregoing intra-model de-idealization, \textit{just like electromagnetic waves}. Begin with electromagnetism. Standard antenna theory tells us that a dipole must sit in the far field before the oscillating electric field can be approximated by a genuine plane wave; only then does the Poynting flux fall like $1/r^2$ and let us read off, with well-controlled errors, the radiative energy absorbed in the load. Move the antenna closer to its source, into the `near-field' region, and the same Poynting vector becomes ambiguous; the wave/no-wave split has blurred and we cannot clearly attribute wave-like behaviour to the Poynting vector.  As Gomes and Rovelli point out, exactly the same fall-off hierarchy appears in GR. Place Feynman's sticky-bead detector far enough from a binary black-hole source such that the curvature peels down to its $1/r$ radiative component in terms of $\Psi_4$, and the predictions are well-known: the beads slide, friction warms the rod, and energy is transferred, quantifiably, to the rod from the gravitational waves. However, when the beads sit too near the source, one cannot cleanly isolate the radiative part, just as an antenna in the near field cannot decide what portion of a nonzero Poynting flux represents true radiation. Yet, we can still measure the latter, so we have good reasons to think that we are measuring the former. Furthermore, they note that this reasoning conforms to the sorts of measurements actually made at LIGO: ``[The] scale separations between the size of our LIGO detectors and the distance to what it is observing suffice for us to identify a component of the gravitational field here that represents the outgoing radiation emitted by the astrophysical sources." \parencite[4]{gomes_rovelli_2024} Thus, the empirical control, and arguments for why both waves are nonetheless measurable, are remarkably similar in physically relevant ways: ``the accounting of energy is explicit, just as it is with electromagnetic radiation." \parencite[9]{gomes_rovelli_2024}  The rod heats for exactly the reason the antenna does; hence, if electromagnetic waves carry energy, gravitational waves must carry it too.

Taken together, the two moves justify speaking of the energy of gravitational waves despite its reliance on strong idealizations. The initial formal bridge between electromagnetic and gravitational waves guarantees mathematical continuity, while de-idealization strategies available to both, controlling both the errors and how these errors are negligible in measurement contexts, provide a strong argument for physical, not merely formal, similarity. \textit{If} we accept their argument for physical similarity, we have a successful inter-model de-idealization across domains: we provide justification for modelling gravitational waves and their status as energy carriers using the idealized spacetime model, in virtue of the conceptual continuity of a \textit{wave} -- both formal and physical similarities -- across idealized gravitational wave models and more realistic electromagnetic wave models. \x{Furthermore, given successful inter-model de-idealization across domains, de-idealization strategies already known to work in the more realistic model can guide refinements of the idealized model as well, potentially leading to further inter-model de-idealization and strengthening the justification already obtained.} 

\section{Measurement de-idealization}
Intra-model and inter-model de-idealizations provide significant justificatory strength in licensing the use of idealizations in physics. However, we recognize, in the spirit of e.g.\ \cite{Knuuttila2019-KNUDNE}, that such de-idealizations are challenging to find. This motivates a third strategy: \textit{measurement de-idealization}. It differs in kind from the other two strategies: rather than closeness to other models in a parameter-family, or to other families of models within/across  domains, \x{the key notion of closeness here is \emph{robust, diachronic convergence} between (i) a history of predictions produced by successive refinements of an idealized model and (ii) a history of observational or experimental measurements of the relevant target properties.}

\x{While a limiting case would be a single successful prediction against a single measurement with no subsequent refinements, this generally provides little by way of de-idealization. Rather, what matters is a sustained pattern: as idealized features of the model are refined, the model's predictions should get closer, by contextually appropriate standards,\footnote{Consider, for instance, particle physics' use of 5-sigma as a gold standard of approximate confirmation.} to an array of measurements obtained across different experimental and observational settings. Note, again, the context-dependence of de-idealization: idealizations are measurement de-idealized relative to a set of measurement contexts (possibly over space and time), and, of course, what is ``close'' depends on the sorts of measurements, sources of errors, and how other available models perform on the same measurements.}

\x{Measurement de-idealization provides justification for the use of an idealized model in three ways. First, at each stage of refinement, the gap between the idealized model's predictions and the measurement outcomes constitutes an error-bound, specifying applicability conditions for the idealized model; the more \textit{robust} this gap is -- the more we consistently observe the gap across a variety of contexts -- the more we are given license to apply the idealized model at that stage of refinement only if the error-bound is negligible for the context of modelling.\footnote{Indeed, modern statistical methods for quantifying error bounds and distinguishing random and systematic error can be seen as a byproduct of measurement de-idealization, dating back to Boscovich and Laplace's analyses of the robustness of gaps between Newton's model of Earth's equilibrium figure and the observed data. See \textcite[\S2]{ohnesorge_2024}.} This distinguishes measurement de-idealization from confirmation: it is not just about what our models get \textit{right}, but also the robustness of what they get \textit{wrong} that provides license for using the idealized model. Second, the more we observe a history of convergence -- \textit{diachronic convergence} -- between predictions and measurements, the more we are justified in adopting the idealized model and its refinements (again, whose domains of applicability are indexed to their associated error-bounds). Finally, this history of robust diachronic convergence under refinement helps us evaluate the refinements themselves. Insofar as convergence is driven by a particular refinement to the idealized model, this, in turn, justifies that refinement. Contrariwise, if convergence is reversed -- if the gap \textit{widens} after some refinement -- this provides justification \textit{against} including said refinement into the model.}

\x{It is helpful to characterize these gaps in the language of \textit{residuals}, following \textcite{Miyake2023}. Residuals are what persists of a complex phenomenon once the contributions of all currently recognized causes -- characterized \textit{in terms of} the idealized model and its accompanying background assumptions (hence why residuals are theoretical items) -- have been accounted for by the model in question.\footnote{\textcite{smith_closing_2014} calls these ``second-order phenomena''.} Put simply, they are what's not accounted for in the model, by the lights of the model. For instance, we can start with observations of planetary motions in their full complexity, ask the extent to which our idealized models account for these motions, and then ask what remains unexplained \parencite[\S5]{Miyake2023}. Crucially, residuals are not merely disagreement between predictions and measurement; they are structured disagreement. For instance, when it comes to planetary motion, we can ask if the model-observation discrepancy arises secularly or periodically. This structure specifies what the model fails to explain -- and the extent to which the model fails to explain -- by the lights of the model itself. Residuals thus provide a finer-grained understanding of the gaps between model and measurement, and can guide which parts of our models we should modify: in the simple example here, if the residual is periodic, one natural refinement would be to introduce a periodic source of error in our model.}

\x{Construed this way, measurement de-idealization proceeds as such: start with an idealized model, compare its predictions to a suitably rich body of measurements, and identify residual patterns. Given the structure of such patterns, modify aspects of the model to test which refinements reduce those residuals and which do not.\footnote{This also invites us to seek better measurements, so as to refine the residuals themselves.} When this process succeeds over time, residuals shrink in magnitude, stabilize across contexts, or become systematically attributable to well-understood sources of error.\footnote{See \textcite[\S4]{BokulichForthcoming-BOKRTA} for a classification of various ways that residues can de-idealize.} This is the sense in which residuals function as de-idealizing constructs: they provide constraints on the space of acceptable refinements, and thereby guide which idealizations can be retained, which can be relaxed, and which must be abandoned. Put in \textcite{smith_closing_2014}'s terms, successful episodes of measurement de-idealization exemplify ``closing the loop'', where modelling and measurement practices co-evolve under the pressure of residuals.\footnote{This is just a first pass; see \textcite{smith_closing_2014} and \textcite{Stan2023-STATXI}. See \textcite{BokulichForthcoming-BOKRTA} for generalizations of the method of residuals to cases where theories change.}}

\x{One example of residual reasoning comes from lunar theory. Newton's law of gravitation, relying on first-order perturbative analysis, predicted a motion of the lunar apsides that was only about half the observed value, a residual that Herschel reports ``became so great a stumbling-block in the way of succeeding geometers, as to shake their faith in the theory of gravity.''\footnote{Herschel (1829, 720), quoted in \textcite[13]{Miyake2023}.} However, Clairaut carried the perturbative analysis to second order, and the additional terms accounted for the residual; in Herschel's words, this turned the anomaly into ``a most cogent argument in [gravity's] favour,'' since the theory now matched not only the leading behaviour but \textit{also} the ``refinements and niceties'' of the data.\footnote{Herschel (1829, 724), quoted in \textcite[13]{Miyake2023}.} This is measurement de-idealization in action in precisely the sense characterized above: the residual provides an error-bound that initially undermines the domain of applicability of the first-order treatment, while the second-order refinement reduced the residual in a structured way. The resulting convergence provides justification both for the refinement and for the continued use of the idealized model within regimes where remaining residuals are negligible.}

\x{While successful refinements are sometimes also intra-model or inter-model de-idealizations in our earlier senses, and those can add further justificatory strength, we emphasize that measurement de-idealization does not require that such an option is available -- the refinements may not be so principled or thorough-going as we'll see below.\footnote{For instance, measurements and residual reasoning might proceed without secure theoretical footing; cf. \textcite{Ohnesorge2023-OHNTEP}. See also \textcite{larrouletphilipppi_lessons_measurement_forthcoming} for similar themes of measurement de-idealization without theoretical footing.} Rather, the core point is that residual structure constrains refinement in ways that are neither arbitrary nor merely pragmatic: it is not the bare fact of agreement -- and confirmation -- that does epistemic work, but the robustness of error-bounds across contexts and the diachronic pattern of which refinements genuinely drive convergence.}

Consider familiar discussions of the Bohr model \parencite{Bokulich2011-BOKHSM} and the Ising model \parencite{Weisberg2007-WEITKO-2}. In both cases, it is argued that these models lack \textit{full} de-idealization. Both Bokulich and Weisberg then consider other pragmatically-driven ways of understanding such models via how they produce explanations, as in the case of minimal models. However, we suggest that measurement de-idealization -- refinement-sensitive diachronic convergence -- did much of the justificatory work that made these idealizations worth adopting and developing in the first place, and that it also explains how their epistemic standing can be strengthened or withdrawn over time.

Consider first the Bohr model. Why did it gain traction as a model \emph{of} hydrogen at all? A key episode was ``the fit between the theoretical and the measured values [of the Rydberg constant]'' which ``was good to three figures, a striking validation of the model'' \parencite[260]{McMullin1985-MCMGI-2}. The Bohr model initially yielded successful predictions (e.g., of hydrogen's emission spectrum) across a variety of contexts, and the residual discrepancies that remained were, at least initially, small enough to be plausibly treated as higher-order effects or experimental limitations. As the model was refined -- by incorporating relativistic corrections, additional quantum numbers, and the characteristic modifications of the Bohr--Sommerfeld program -- physicists could track which changes reduced residuals and which failed. \x{This provided exactly the three forms of justification described above: (i) an evolving error-bound that delimited when the idealized model could be safely applied, (ii) a diachronic pattern of convergence that supported continued reliance within a circumscribed domain, and (iii) refinement-sensitive constraints that helped discriminate which structural elements were doing genuine work in the model's successes.}\footnote{See \textcite[table 8.1, p.\ 347]{kragh2012niels} for a list of the Bohr model's empirical successes and failures.} As \textcite[p.\ 90]{kragh2012niels} emphasizes, ``what swayed otherwise sceptical physicists to accept it was primarily its empirical successes, that is, its remarkable ability to account for or predict phenomena that other theories failed to explain.'' 

Just as importantly, the Bohr case also illustrates how measurement de-idealization can fail when persistent residuals resist refinement. The convergence that initially supported it broke down as new experimental and measurement contexts emerged. Kragh notes a major reason why physicists abandoned the Bohr-Sommerfeld model (even \textit{prior} to Schr\"odinger's model and the `new' quantum theory): ``persistent discrepancies between the Bohr-Sommerfeld theory and experimental results,'' that is, residuals which could not be eliminated even after many iterations of intra-model de-idealizations of the original Bohr model -- justification broke down in these new contexts.\footnote{\x{Note, however, that we can -- and do -- continue to use the Bohr-Sommerfeld model as a first approximation (informed by its error-bound in successful contexts), especially for understanding the quantization of energy levels and ground states in the contexts where measurement de-idealization succeeded, e.g., for the hydrogen atom.}} Divergence between the Bohr model and measurements signaled a failure to close the loop. Then, even within the original domains, e.g.\ the hydrogen atom, where the Bohr model found success, the predictions of Schr\"odinger's model better approximated measurement outcomes, thereby providing stronger justification for adopting the new quantum theory over retaining Bohr's model.  

Compare this with the Ising model, for which we might ask again: \textit{why} did scientists deem it worth studying, such that they then tried to develop a rich story for how the highly idealized model maps onto the `privileged causal factors' \cite[p. 645]{Weisberg2007-WEITKO-2} of real-world phenomena? In particular, what reasons did scientists have for expecting explanations provided by the Ising model to be those \textit{of} real-world phenomena to begin with, without intra-/inter-model de-idealizations?  

It is important to note that the Ising model was \textit{not} accepted as a useful model for studying real-world phenomena when it was first developed in 1920 by Lenz and Ising. Instead, as \textcite[251]{Niss2008-NISHOT-2} describes, most physicists initially took the model to be distorting the behaviour of magnetic spin and was simply irrelevant to the study of real-world magnets. So even if it \textit{could} be seen as highlighting some causal factors, it was doing it \textit{inaccurately} and did not approximate experimental observations; residuals persisted and there was no robust convergence -- no measurement de-idealization. \textcite[48]{benedek1966a}, in a study of experimental data from 1961 to 1965, for instance, concluded that ``by and large the experimental results indicate that the Ising model is inadequate to describe a real ferromagnet.'' 

Even the widely touted use of Ising models for critical phenomena did not come easy. As C.N. Yang noted, Onsager's 1944 development of Ising models for describing critical phenomena was initially deemed ``a mathematical curiosity with no real physical relevance" and people who studied the Ising model were described as ``contracting the Ising disease." \parencite[3]{yang1972a} Why, then, did it eventually gain justification for modelling critical phenomena in the 1960s? As \textcite[281]{Niss2008-NISHOT-2} points out, newfound ``experimental agreement with [the Ising model] played an important role in recognizing its significance". \cite{yang1972a} observed that the initial scepticism ``disappeared during the 1960s when it became clear that the lattice gas description of liquid-gas transitions captured much of the essential features of the singularities'' found in new experimental measurements.\footnote{These measurements revealed thermodynamic quantities becoming singular at transition points.} Notably, it did \textit{better}, in comparison to classical mean-field and phenomenological models, in predicting the critical exponent $\delta$.\footnote{$\delta$ features in the critical isotherm, the curve of pressure $P$ as a function of volume $V$ at constant critical temperature $T_C$: $(P - P_C) \sim (V - V_C)^\delta$.} While the latter models predicted $\delta = 3$, \cite{gaunt1964a}'s Ising model predicted $\delta = 5.20 \pm 0.15$. The experimental data suggested a value of $\delta \approx 4.2$. Despite large discrepancies, the Ising model nonetheless better converged -- diachronically and robustly -- to the measured quantity than its competitors. This provided support for the Ising model over then-existing classical models. As \textcite[947]{fisher1964a} observes, such convergence ``suggests that in the critical region the lattice gas represents rather adequately the pertinent features of a real gas."

Furthermore, unlike the Bohr model, convergence persisted in new experimental contexts: subsequent refinements of the Ising model and experimental setups led to \textit{improved} convergence -- rather than divergence -- between the predicted and measured quantity of $\delta$, as well as critical exponents and specific heats in gas-liquid transitions among other things \parencite[277-280]{Niss2008-NISHOT-2}, shrinking residuals and providing continued and increased justification both to use the Ising model as a model \textit{of} critical phenomena and to incorporate said refinements into the model.

Some like McMullin might denigrate measurement de-idealization as ``merely'' fitting the data or ``saving the appearances'' (\citeyear[262]{McMullin1985-MCMGI-2}). 
Notably, he attacks quantum field theory's (QFT) use of (old) renormalization techniques -- ``techniques for which no theoretical justification can be given'' -- for fixing divergent quantities in QFT: ``The model itself in such a case is suspect, \textit{no matter how good the predictive results it may produce}. Scientists will work either to derive the corrections theoretically, or else to replace the model with a more coherent one.'' (\citeyear[261]{McMullin1985-MCMGI-2}, emphasis ours.)

To us, this critique is incomplete. On our view, an idealized model can earn substantial \textit{pro tanto} justification from measurement de-idealization even when it lacks intra-/inter-model de-idealization. It is \textit{not} easy for any particular idealized model to `save the appearances' in the robust diachronic sense we have been considering, especially for the increasingly complicated phenomena we find in science -- the Bohr model failed to do so even after iterations of refinements. That is, models that `save the appearances' turn out to be remarkably constrained by the world; more so, a model which does so to incredible levels of accuracy over an extremely wide range of experimental contexts -- such as quantum electrodynamics (QED) in QFT.

\x{QED's early history makes this vivid. When \textcite{Dyson1949} extended QED by introducing the methods of perturbation theory, he was explicit that the resulting refined framework was, at that stage, to be vindicated via its applications -- in generating fruitful predictions -- rather than treated as a ``theoretical derivation'':\footnote{~``...the theory as a whole cannot be put into a finally satisfactory form so long as divergencies occur in it, however skillfully these divergencies are circumvented... the present treatment should be regarded as justified by its success in applications rather than by its theoretical derivation.'' \parencite[490]{Dyson1949}} the series is used without a general proof of convergence, no assurance that blow-ups will not occur at higher-order calculations -- without such a proof, there is no intra-model de-idealization. Yet, it seems wrong to say that the Dyson series was unjustified simpliciter: the refined framework recovered not only earlier QED predictions and observations, but predicted new phenomena that further reduced the residuals (e.g. higher order radiative reactions and vacuum polarization phenomena). Of course, later in the 1970s, Wilsonian renormalization-group techniques explained why such procedures worked -- providing intra-model de-idealization -- but note that this came much later. We see no reason to say that QED -- and refinement via the Dyson series -- was only justified then. We grant that this justification is pro tanto, but we believe that intra-model de-idealization only added further justification on top of measurement de-idealization. Contra McMullin, we see a role for `saving the appearances' if this salvation comes in robust, refinement-sensitive, diachronic ways.}

\section{Conclusion}

When it comes to de-idealization, to aim at reversing to a ``full representation" is itself a philosophically loaded idealization. What physicists actually do, as we hope to have shown, is to forge many smaller, context-specific bridges that keep an idealized model in contact with the world. We have highlighted three such bridges for de-idealization: intra-model, inter-model, and measurement de-idealization. None requires the mythical ``full representation"; all are rooted in practices in physics. 

The skeptic may of course demur. On one extreme, one can insist that only full de-idealization counts as de-idealization, because only truth matters, not more realistic models which may be idealized in some other ways. On the other extreme, one can insist that modelling has nothing to do with truth, and everything is mere pragmatics. We seek middle ground. \y{For us, de-idealization is an ongoing epistemic enterprise: a sustained effort to show that idealized models remain appropriately constrained by, and responsive to, the world -- without requiring that they be fully or literally true.} This is in line with a recent trend in philosophy of science, which places less emphasis on synchronic analyses of how theories and models relate to the world at a time -- or once and for all -- from an external standpoint that presupposes access to the ``real world" or to full representations. In practice, all we have are idealized models, and we can only work from within them. Seen this way, de-idealization is an ongoing \textit{process}. \y{We are constantly in search of better de-idealizations -- ways of justifying how our idealized models support reliable inferences about the world. On the flipside, this means that our models are never fully justified, once and for all -- they are not full and exact representations!} Rather, we are constantly in the process of `true-ing' our idealized models. As \textcite[3]{Andersen2023-ANDT-3} puts it, true-ing ``emphasizes the process by which models are brought into true with their target systems, and the fact that this process is not accomplished once and then is done forever, but instead requires upkeep and ongoing fine-tuning even in the best of cases." Intra-model, inter-model, and measurement de-idealizations are strategies for true-ing our models over time. They are not mutually exclusive strategies: successful measurement de-idealization can lead to more work on the model, and to the search for intra-model or inter-model de-idealizations, leading to higher and higher justificatory strength for idealized models, \y{helping us check our models by showing in what respects they reliably inform our understanding of the world.} Thus, rather than leaving idealizations ``unchecked'', we would say: ``Check, please!" 

\numberwithin{equation}{section}
\numberwithin{figure}{section}
\numberwithin{table}{section}
\section*{Acknowledgments}
 
We would like to thank Holly Andersen, Zili Dong, Hanti Lin, Chia-Hua Lin, Teru Miyake, Wayne Myrvold, Sai Ying Ng, Miguel Ohnesorge, Angela Potochnik, Carlo Rovelli, Chris Smeenk, Jacob Stegenga, Francesca Vidotto, Shimin Zhao, attendees of the UIC Undergraduate Philosophy Club meeting, attendees of the UCSD Graduate Philosophy Colloquium, attendees of the Rotman Graduate Student Conference (RGSC), and attendees of the Pragmatism and Philosophy of Science workshop at the University of Western Ontario, for their comments and suggestions. We would also like to thank two anonymous referees for their comments and suggestions throughout the review process.

\newpage
\begin{flushright}

  Yichen Luo \\
  \emph{School of Humanities \\ 
  Nanyang Technological University\\
  Singapore, Singapore} \\
  and \\
  \emph{Rotman Institute of Philosophy\\
  The University of Western Ontario\\
  London, Ontario, Canada\\
  yichenaluo@gmail.com
}
\end{flushright}
\begin{flushright}

  Eugene Y. S. Chua \\
  \emph{School of Humanities\\ 
  Nanyang Technological University\\
  Singapore, Singapore \\
  eugene.chuays@ntu.edu.sg 
}
\end{flushright}
\printbibliography[title ={References}]

\end{document}